\def\ra{\rangle}
\def\la{\langle}
\def\be{\begin{equation}}
\def\ee{\end{equation}}
\def\ba{\begin{array}}
\def\ea{\end{array}}

\documentclass[aps,pre,preprint,showpacs,amsmath,amssymb,amsfonts,preprintnumbers]
{revtex4}
\usepackage{epsfig,graphicx}

\begin{document}

\baselineskip=18pt \setcounter{page}{1} \centerline{\large\bf
Measurable bounds for Entanglement of Formation } \vspace{4ex}
\begin{center}
Ming Li$^{1}$, Shao-Ming Fei$^{2,3}$

\vspace{2ex}

\begin{minipage}{5in}

{\small $~^{1}$  College of Mathematics and Computational Science,
China University of Petroleum, 257061 Dongying}

{\small $~^{2}$ Department of Mathematics, Capital Normal
University, 100037 Beijing}

{\small $~^{3}$ Max-Planck-Institute for Mathematics in the Sciences, 04103 Leipzig}

\end{minipage}
\end{center}

\begin{center}
\begin{minipage}{5in}
\vspace{1ex} \centerline{\large Abstract} \vspace{1ex}
We study the entanglement of formation for arbitrary dimensional
bipartite mixed unknown states. Experimentally measurable
lower and upper bounds for entanglement of formation are derived.

\smallskip
PACS numbers: 03.67.-a, 02.20.Hj, 03.65.-w\vfill
\smallskip
\end{minipage}\end{center}
\bigskip

Being one of the most striking phenomena in quantum physics, quantum
entanglement \cite{ent1,ent2,ent3} has been extensively investigated
in recent years. One of the main tasks in quantum entanglement
theory is to quantify the entanglement of
quantum systems. Among all the bipartite entanglement measures,
entanglement of formation (EOF) is one of the most meaningful and
physically motivated measures
\cite{BDSW,Horo-Bruss-Plenioreviews}, which quantifies the minimal
cost needed to prepare a certain quantum state in terms of EPR
pairs, and plays important roles in many physical systems, such as quantum phase
transition for various interacting quantum many-body systems
\cite{Osterloh02-Wu04}, macroscopic
properties of solids \cite{Ghosh2003}, and capacity of quantum channels \cite{Shor-Pomeransky}.

Let $H_A$, $H_B$ be the $m$, $n$ ($m\leq n$) dimensional vector spaces respectively.
A pure quantum state $|\psi\ra\in H_A \otimes H_B$ is an $mn$-dimensional vector. Its
entanglement of formation is defined by $E(|\psi\ra)=S(\rho_A)$,
where $\rho_A=Tr_B(|\psi\ra\la\psi|)$ is the reduced density matrix of $|\psi\ra\la\psi|$.
$S(\rho_A)$ is the entropy
\begin{eqnarray}
S(\rho_A)=-\sum_{i=1}^m\mu_i \log\mu_i\equiv H(\vec{\mu}),
\end{eqnarray}
where $\log$ stands for the natural logarithm throughout the paper,
$\mu_i$ are the eigenvalues of $\rho_A$ and $\vec{\mu}$ is the
Schmidt vector $(\mu_1,\mu_2,\cdots,\mu_m)$. This definition of
entanglement of formation is extended to mixed states $\rho$ by the
convex roof,
\begin{eqnarray}
E(\rho)=\min_{\{p_i,|\psi_i\ra\}}\sum_ip_iE(|\psi_i\ra),
\end{eqnarray}
for all possible ensemble realizations
$\rho=\sum_ip_i|\psi_i\ra\la\psi_i|$, where $p_i\geq 0$ and
$\sum_ip_i=1$.

It is a great challenge to find an analytical formula of the
entanglement of formation for general bipartite quantum mixed states
$\rho$. Considerable efforts have been made on deriving entanglement
of formation or its lower bound through analytical and numerical
approaches. So far the entanglement of formation has been calculated
for some particular states like bipartite qubit states \cite{qubit},
isotropic states \cite{2625} and Werner states \cite{werner} in
arbitrary dimensions, and symmetric gaussian states in infinite
dimensions \cite{gaus}. In \cite{kai} in order to estimate the
entanglement of formation for general states, a lower bound has been
presented by using the partial transposition of $\rho$ with respect
to the subsystem $H_A$, $\rho^{T_{A}}$, and  the realignment of
$\rho$, $R(\rho)$. It is shown that (without regard to the
normalization coefficient $\log 2$) \be
E(\rho)\geq\left\{\begin{array}{l}
0,~ \Omega = 1;\\
H_2[\gamma(\Omega)]+[1-\gamma(\Omega)]\log_2(m-1),~ 1< \Omega \leq \frac{4(m-1)}{m};\\
\frac{\log_2(m-1)}{m-2}(\Omega-m)+\log_2m,~ \frac{4(m-1)}{m}<\Omega \leq m;
\end{array}\right.
\ee
where $\Omega=\max\{||\rho^{T_{A}}||,||R(\rho)||\}$,
$H_2$ is the standard binary entropy function, $||A||$ denotes the trace norm of the matrix $A$.

Comparing with the entanglement of formation, the entanglement measure concurrence is relatively
easier to be dealt with. In \cite{cafc} a simpler analytical lower bound for
concurrence has been presented. And a series of new results related to the bounds of concurrence
have been further obtained \cite{lbc}. In particular
in \cite{mintert, zhang, aolita}, the authors
derive measurable lower and upper bounds for concurrence for general mixed quantum states,
\be\label{lbc2}
2[{\rm Tr}\rho^2-{\rm
Tr}\rho_A^2]={\rm Tr}(\rho\otimes\rho V_i)\leq[C(\rho)]^2\leq {\rm
Tr}(\rho\otimes\rho K_i)=2[1-{\rm Tr}\rho_A^2],
\ee
where $V_1=4(P_--P_+)\otimes P_-, V_2=4P_-\otimes(P_--P_+)$ and
$K_1=4P_-\otimes I$, $K_2=4I\otimes P_-$, $P_-$ is the projector on the antisymmetric subspace of the
two copies of either subsystem, $P_+$ the symmetric counterpart of $P_-$.

Contrary to the concurrence, less has been achieved related to the lower and upper bounds
of entanglement of formation.
In this paper, we derive analytical lower and upper bounds for
entanglement of formation which care measurable experimentally. These bounds supply
nice estimation for entanglement of formation for some quantum states.

To derive lower and upper bounds for entanglement of formation, we first consider a pure
$|\psi\ra$ with Schmidt decomposition
$|\psi\ra=\sum_{i=1}^m\sqrt{\mu_i}|ii\ra$, where $\mu_i\geq 0,
\sum_i^m\mu_i=1$. It is easily verified that
\be
1-{\rm Tr}\rho_A^2={\rm Tr}\rho^2-{\rm Tr}\rho_A^2=1-\sum_i\mu_i^2\equiv
\lambda.
\ee
Set
\be
X(\lambda)=\max\{H(\vec{\mu})|1-\sum_i\mu_i^2\equiv \lambda\},\ \
Y(\lambda)=\min\{H(\vec{\mu})|1-\sum_i\mu_i^2\equiv \lambda\}.
\ee
Let $\varepsilon(x)$ be the largest convex function that is
bounded above by $Y(x)$ and $\eta(x)$ the smallest concave
function that is bounded below by $X(x)$.

{\bf{Theorem:}} For any $m\otimes n (m\leq n)$ quantum state $\rho$,
the entanglement of formation $E(\rho)$ satisfies
\be\label{th}
\max\{\varepsilon({\rm Tr}\rho^2-{\rm Tr}\rho_A^2),\varepsilon({\rm
Tr}\rho^2-{\rm Tr}\rho_B^2)\}\leq E(\rho)\leq \min\{\eta(1-{\rm
Tr}\rho_A^2),\eta(1-{\rm Tr}\rho_B^2)\}.
\ee

{\bf{Proof:}} Without lose of generality, we assume that
$\varepsilon({\rm Tr}\rho^2-{\rm Tr}\rho_A^2)\geq \varepsilon({\rm
Tr}\rho^2-{\rm Tr}\rho_B^2)$ and $\eta(1-{\rm
Tr}\rho_A^2)\leq\eta(1-{\rm Tr}\rho_B^2)$.
Note that for any pure state $|\psi\rangle$, the concurrence is given by
$C(|\psi\rangle)=\sqrt{2(1-{\rm Tr}(\rho^A)^2)}$.
Due to convexity of concurrence, for any pure
decomposition $\rho=\sum_\alpha p_\alpha \rho_\alpha$, we have
$$
\sum_{\alpha}p_{\alpha}C^2(\rho_\alpha)=
\sum_{\alpha}2\,p_{\alpha}[1-{\rm Tr}(\rho_{\alpha}^A)^2]\geq
C^2(\rho).
$$
Taking into account of the bounds (\ref{lbc2}) of $C^2(\rho)$, we obtain
\be\label{z1}
\sum_\alpha p_\alpha
[{\rm Tr}(\rho_\alpha)^2-{\rm Tr}(\rho^A_\alpha)^2]\geq {\rm
Tr}\rho^2-{\rm Tr}\rho_A^2
\ee
and
\be\label{z2} \sum_\alpha p_\alpha
[1-{\rm Tr}(\rho^A_\alpha)^2]\leq 1-{\rm Tr}\rho_A^2.
\ee

Assume $\rho=\sum_\alpha p_\alpha \rho_\alpha$ be the optimal
decomposition of $E(\rho)$. We have \be E(\rho)=\sum_\alpha p_\alpha
E(|\psi_\alpha\ra)=\sum_\alpha p_\alpha
H(\vec{\mu}_\alpha)\geq\sum_\alpha p_\alpha
\varepsilon(\lambda_\alpha)\geq \varepsilon(\sum_\alpha
p_\alpha\lambda_\alpha)\geq\varepsilon(\Lambda), \ee where
$\Lambda={\rm Tr}\rho^2-{\rm Tr}\rho_A^2$. We have used the
definition of $\varepsilon$ to obtain the first inequality. The
second inequality is due to the convex property of $\varepsilon(x)$
and the last one is derived from (\ref{z1}).

On the other hand, \be E(\rho)=\sum_\alpha p_\alpha
E(|\psi_\alpha\ra)=\sum_\alpha p_\alpha
H(\vec{\mu}_\alpha)\leq\sum_\alpha p_\alpha \eta(\lambda_\alpha)\leq
\eta(\sum_\alpha p_\alpha\lambda_\alpha)\leq\eta(\Lambda^{'}), \ee
where $\Lambda^{'}=1-{\rm Tr}\rho_A^2$. We have used the definition
of $\eta$ to get the first inequality. The second inequality is
derived from the concave property of $\eta(x)$ and the last one is
obtained from (\ref{z2}). $\hfill\Box$

We calculate now both the maximal admissible $H(\vec{\mu})$ and the
minimal admissible $H(\vec{\mu})$ for a given $\lambda$, i.e.
$X(\lambda)$ and $Y(\lambda)$, by using the Lagrange multipliers approach
\cite{2625}. The necessary conditions for the maximum and minimum are given by:
\be\label{e12}
-\log\mu_k-1-2x\mu_k+y=0,
\ee
\be\label{eqs}
1-\sum_i\mu_i^2-\Lambda=0,\ \ 1-\sum_i\mu_i=0;
\ee
where $x, y$ denote the Lagrange multipliers. We get from (\ref{e12}) that
\be
\label{lag} \log\mu_k=-2x\mu_k+y-1.
\ee
From (\ref{lag}) we know
that there are at most two solutions for each $\mu_k$, which will be
denoted as $\alpha$ and $\beta$ in the following.

Let $n_1$ be the number of entries where $\mu_i=\alpha$ and $n_2$
the number of entries where $\mu_i=\beta$. If one of the $n_1$ and $n_2$ is
zero, we have that $\lambda=1-\frac{1}{n_1+n_2}$. Otherwise
both $\alpha$ and $\beta$ are nonzero. The problem is now turned to be,
for fixed $n_1, n_2, n_1+n_2\leq m$, one
maximizes or minimizes the function
\be\label{14}
F_{n_1n_2}(\lambda)=n_1 h(\alpha)+ n_2 h(\beta),
\ee
where $h(x)=-x\log x$, under the constraints (\ref{eqs}). By direct
computation we obtain
\be
\alpha_{n_1n_2}^{\pm}=\frac{n_1\pm\sqrt{n_1^2-n_1(n_1+n_2)[1-n_2(1-\lambda)]}}{n_1(n_1+n_2)},
~~~ \beta_{n_1n_2}^{\pm}=\frac{1-n_1\alpha^{\pm}}{n_2}.
\ee
To ensure the nonnegativity property of $\alpha_{n_1n_2}^{\pm}$ and
$\beta_{n_1n_2}^{\pm}$, we require that $\max\{1-\frac{1}{n_1},
1-\frac{1}{n_2}\}\leq\Lambda\leq 1-\frac{1}{n_1+n_2}$. Since
$\alpha_{n_2n_1}^{-}=\beta_{n_1n_2}^+,
\beta_{n_2n_1}^-=\alpha_{n_1n_2}^{+}$, the function in Eq.
(\ref{14}) takes the same value for $\alpha_{n_1n_2}^{+}$ and
$\alpha_{n_2n_1}^{-}$. Therefore we can restrict ourselves to the
solutions $\alpha_{n_1n_2}=\alpha_{n_1n_2}^{+}$. Eq. (\ref{14})
then turns out to be
\be\label{17}
F_{_{n_1n_2}}(\lambda)=n_1 h(\alpha_{n_1n_2}^{+})+ n_2 h(\beta_{n_1n_2}^{+}).
\ee

When $m=3$, to find the expressions of upper and lower bounds in (\ref{th}) is to obtain
the max- and minimization over the
three functions $F_{12}(\Lambda)$, $F_{21}(\Lambda)$ and $F_{11}(\Lambda)$.
From (\ref{17}) for $m=3$ we have
\be\label{18}
X(\Lambda)=\left\{\begin{array}{l}
F_{11},~~ 0< \Lambda\leq \frac{1}{2};\\
F_{12},~~ \frac{1}{2}< \Lambda\leq \frac{2}{3}
\end{array}\right.~~~
Y(\Lambda)=\left\{\begin{array}{l}
F_{11},~~ 0< \Lambda\leq \frac{1}{2};\\
F_{21},~~ \frac{1}{2}< \Lambda\leq \frac{2}{3}.
\end{array}\right.
\ee From (\ref{18}) we have that $\eta[\Lambda]$ is the broken line
connecting the following points:
$[0,0],[\frac{1}{2},\log{2}],[\frac{2}{3},\log{3}].$

\begin{figure}[tbp]
\begin{center}
\resizebox{10cm}{!}{\includegraphics{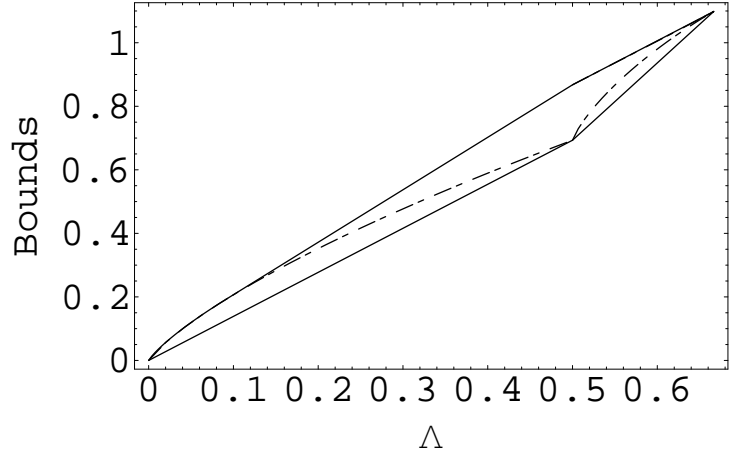}}
\end{center}
\caption{Upper and lower bounds of $E(\rho)$ (solid lines) for
$m=3$, and $F_{11}, F_{12}, F_{21}$(dashed line). \label{fig1}}
\end{figure}

In order to determine $\varepsilon[\Lambda]$ we solve the following
equations: Let $l(\Lambda)=k(\Lambda-0.5)+0.868$ be the line
crossing through the point $[0.5,F_{12}(0.5)]$. We solve (i)
$l(\Lambda)=F_{11}$ and (ii)
$\frac{dl(\Lambda)}{d\Lambda}=k=\frac{dF_{11}(\Lambda)}{d\Lambda}$
for $k$ and $\Lambda$, and find the values to be $1.65$ and $0.091$.
Thus we derive that $\varepsilon[\Lambda]$ is the curve consisted of
$F_{11}$ for $0< \Lambda\leq 0.091$ and a broken line connecting
points $[0.091,F_{11}(0.091)],[0.5,F_{12}(0.5)]$ and
$[0.667,\log[3]]$, i.e. \be\label{19}
\eta[\Lambda]=\left\{\begin{array}{l}
2\log{2}\times\Lambda,~~ 0< \Lambda\leq 0.5;\\
6\log{\frac{3}{2}}\times(\Lambda-\frac{1}{2})+\log{2},~~ 0.5<
\Lambda\leq 0.667
\end{array}\right.
\ee
and
\be
\varepsilon[\Lambda]=\left\{\begin{array}{l}\label{20}
F_{11},~~ 0< \Lambda\leq 0.091;\\
1.65(\Lambda-0.5)+0.868,~~ 0.091< \Lambda\leq 0.5;\\
1.39(\Lambda-0.667)+1.099,~~ 0.5< \Lambda\leq 0.667,
\end{array}\right.
\ee
see Fig.\ref{fig1}.

\begin{figure}[h]
\begin{center}
\resizebox{10cm}{!}{\includegraphics{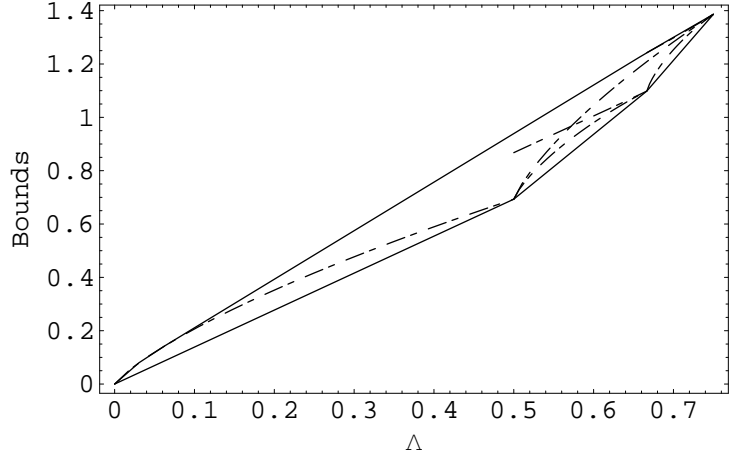}}
\end{center}
\caption{Upper and lower bounds of $E(\rho)$ (solid lines) for
$m=4$, and $F_{11}, F_{12}, F_{21}, $ $F_{22}, F_{13},
F_{31}$(dashed line). \label{fig2}}
\end{figure}

When $m=4$, we need to find the max- and minimization over the
six functions $F_{11},F_{12}, F_{21},F_{22},F_{31},$ and $F_{13},$
which are plotted in Fig.\ref{fig2}. We have
\be
X(\Lambda)=\left\{\begin{array}{l}
F_{11},~~ 0< \Lambda\leq \frac{1}{2};\\
F_{12},~~ \frac{1}{2}< \Lambda\leq \frac{2}{3};\\
F_{13},~~ \frac{2}{3}< \Lambda\leq \frac{3}{4}
\end{array}\right.~~~
Y(\Lambda)=\left\{\begin{array}{l}
F_{11},~~ 0< \Lambda\leq \frac{1}{2};\\
F_{21},~~ \frac{1}{2}< \Lambda\leq \frac{2}{3};\\
F_{31},~~ \frac{2}{3}< \Lambda\leq \frac{3}{4}.
\end{array}\right.
\ee

Further more, one obtains that $\eta[\Lambda]$ is the broken line
connecting the following points:
$[0,0],[\frac{1}{2},\log{2}],[\frac{2}{3},\log{3}],[\frac{3}{4},\log{4}]$
and $\varepsilon[\Lambda]$ is the curve consisted of $F_{11}$ for
$0< \Lambda\leq 0.062$ and a broken line connecting points
$[0.062,0.142],[0.667,1.242]$ and $[0.75,1.386]$(These points can be
derived by using the same processes as that have been done in case
$m=3$), i.e. \be \eta[\Lambda]=\left\{\begin{array}{l}
2\log{2}\times\Lambda,~~ 0< \Lambda\leq 0.5;\\
6\log{\frac{3}{2}}\times(\Lambda-\frac{1}{2})+\log{2},~~ 0.5<
\Lambda\leq
0.667;\\
12\log{\frac{4}{3}}\times(\Lambda-\frac{2}{3})+\log{3},~~ 0.667<
\Lambda\leq 0.75
\end{array}\right.
\ee
and
\be \varepsilon[\Lambda]=\left\{\begin{array}{l}
F_{11},~~ 0< \Lambda\leq 0.062;\\
1.820(\Lambda-0.667)+1.242,~~ 0.062< \Lambda\leq 0.667;\\
1.726(\Lambda-0.667)+1.242,~~ 0.667< \Lambda\leq 0.75.
\end{array}\right.
\ee

Generally, we have the following observation,
for any $m$,
\be
X(\Lambda)=\left\{\begin{array}{l}
F_{11},~~ 0< \Lambda\leq \frac{1}{2};\\
F_{12},~~ \frac{1}{2}< \Lambda\leq \frac{2}{3};\\
\cdots\\
F_{1(m-1)},~~ \frac{m-2}{m-1}< \Lambda\leq \frac{m-1}{m};
\end{array}\right.\quad\quad
Y(\Lambda)=\left\{\begin{array}{l}
F_{11},~~ 0< \Lambda\leq \frac{1}{2};\\
F_{21},~~ \frac{1}{2}< \Lambda\leq \frac{2}{3};\\
\cdots\\
F_{(m-1)1},~~ \frac{m-2}{m-1}< \Lambda\leq \frac{m-1}{m};
\end{array}\right.
\ee
and
$\eta[\Lambda]$ is the broken line connecting the following
points: $[\frac{i}{i+1},\log{(i+1)}], 0\leq i\leq m-1,$ i.e.
\be\label{2}
\eta[\Lambda]=k(k+1)\log{\frac{k+1}{k}}(\Lambda-\frac{k-1}{k})+\log{k},\ee
for $(k-1) < \Lambda\leq k$ and $k=1,2,\cdots,m-1$. The
representation of $\varepsilon[\Lambda]$ can be also figured out
numerically.

The measurable upper and lower bounds can be used to estimate
the entanglement of formation for an unknown quantum mixed state experimentally.
Consider the following mixed quantum state
\be
\rho=\frac{x}{9}I+(1-x)|\psi\ra\la\psi|.
\ee
where $|\psi\ra=(a,0,0,0,\frac{1}{\sqrt{3}},0,0,0,\frac{1}{\sqrt{3}})^t/\sqrt{{\rm Tr}\{|\psi\ra\la\psi|\}}$.
For $x=0.1$, one has
\be\Lambda={\rm
Tr\{\rho^2\}}-{\rm Tr\{\rho_A^2\}}={\rm Tr\{\rho^2\}}-{\rm
Tr\{\rho_B^2\}}=\frac{1.45+9.21a^2-0.38a^4}{(2+3a^2)^2}\ee
\be\Lambda^{'}=1-{\rm Tr\{\rho_A^2\}}=1-{\rm
Tr\{\rho_B^2\}}=\frac{1.14(0.19+a^2)(9.67+a^2)}{(2+3a^2)^2}.
\ee
Substituting $\Lambda$ and $\Lambda^{'}$ above
into (\ref{19}) and (\ref{20}) respectively, we have the upper and
lower bounds (\ref{th}) for $E(\rho)$, see Fig. 3.

For $x=0.001$, one has
\be\Lambda={\rm
Tr\{\rho^2\}}-{\rm Tr\{\rho_A^2\}}={\rm Tr\{\rho^2\}}-{\rm
Tr\{\rho_B^2\}}=\frac{1.99+11.97a^2-0.004a^4}{(2+3a^2)^2}\ee
\be\Lambda^{'}=1-{\rm Tr\{\rho_A^2\}}=1-{\rm
Tr\{\rho_B^2\}}=\frac{0.01(0.17+a^2)(999.67+a^2)}{(2+3a^2)^2}.\ee
The corresponding bounds of $E(\rho)$ is shown in Fig. 4.
We see that the lower and upper bounds are
closer. And the value for $E(\rho)$ can be estimated
more precisely.

\begin{figure}[h]
\begin{center}
\resizebox{10cm}{!}{\includegraphics{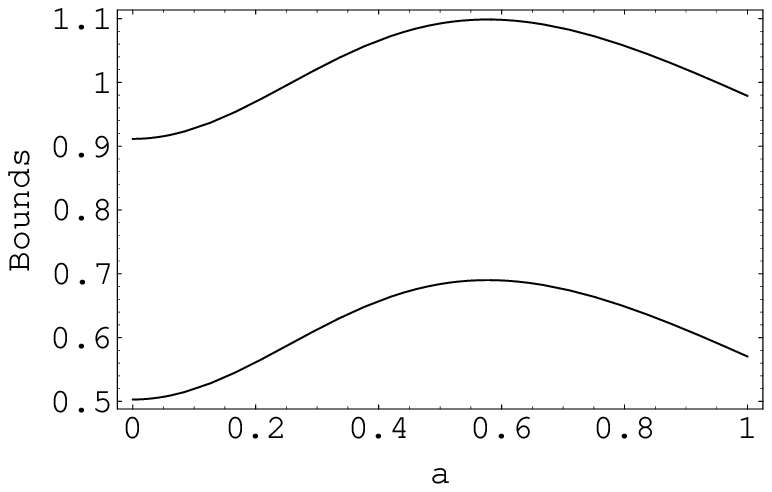}}
\end{center}
\caption{Upper and lower bounds of $E(\rho)$ with $x=0.1$.
\label{fig3}}
\end{figure}

\begin{figure}[h]
\begin{center}
\resizebox{10cm}{!}{\includegraphics{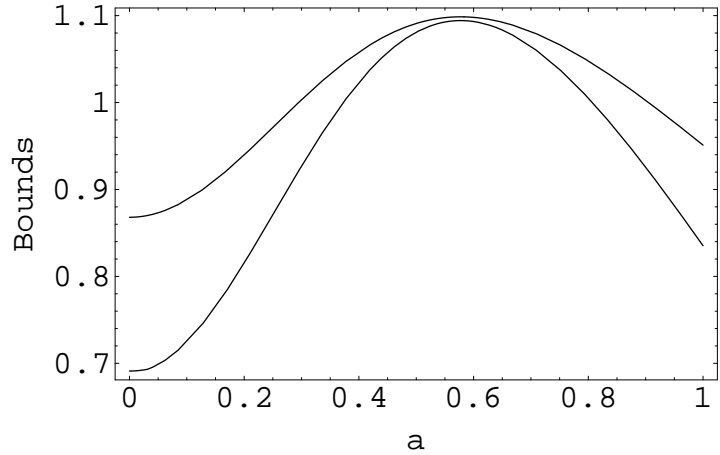}}
\end{center}
\caption{Upper and lower bounds of $E(\rho)$ with $x=0.001$.
\label{fig5}}
\end{figure}

We have studied the entanglement of formation for mixed
quantum states. We have derived upper and lower bounds for entanglement of formation
that are experimentally
measurable. These bounds together can be used to estimate the entanglement of formation
for arbitrary finite dimensional unknown states according to a few
measurements on a twofold copy $\rho\otimes\rho$ of the mixed states.
These results supplement further the estimation for entanglement of formation, like the
case of concurrence for which many lower and upper bounds have been already obtained.

\bigskip
\noindent{\bf Acknowledgments}\, This work is supported by the NSFC 10875081,
KZ200810028013, and PHR201007107.

\end{document}